\documentclass[twocolumn,showpacs,preprintnumbers,amsmath,amssymb]{revtex4-1}

\usepackage{amsmath}
\usepackage{amssymb}
\usepackage{epsfig}
\usepackage{epsf,afterpage}
\usepackage{amssymb}
\newcommand{\be}{\begin{equation}}
\newcommand{\ee}{\end{equation}}
\newcommand{\lan}{\langle}
\newcommand{\rrr}{\rangle}

\newcommand{\T}{\mbox{Tr}\> }
\usepackage{graphicx}
\usepackage{dcolumn}
\usepackage{bm}

\begin{document}

\title{Towards a quantum theory of chiral magnetic effect}

\author{V.D.Orlovsky, V.I.Shevchenko}
\affiliation{%
Institute of Theoretical and Experimental Physics
\\B.Cheremushkinskaya 25, 117218 Moscow, Russia
}%

\date{\today}

\begin{abstract}

We discuss three possible ways of addressing quantum physics behind chiral magnetic effect and electric charge fluctuation patterns
in heavy ion collisions. The  first one makes use of P-parity violation probed by local order parameters, the second considers CME in quantum measurement theory framework and the third way is to study P-odd * P-odd contributions to P-even observables. In the latter approach relevant form-factor is constructed and computed for weak magnetic field in confinement region and for free quarks in strong field. It is shown that the effect is negligible in the former case. We also discuss saturation effect - charge fluctuation asymmetry for free fermions reaches constant value at asymptotically large fields.

\end{abstract}

\pacs{12.38.Aw, 12.38.Mh}

\maketitle

\section{Introduction}

One of the main theoretical challenges of modern quantum chromodynamics (QCD) is to build a detailed theoretical picture of
strong interaction physics relevant for heavy ion collisions. Currently running experimental programs have already brought lots of
exciting results. Despite tremendous progress in understanding, rich pattern of observed effects is still
waiting for being placed into coherent theoretical picture based on QCD.

In the course of studies of hadronic matter at large temperatures
and/or densities one can make use of the scale separation allowing
to neglect effects of weak and electromagnetic interactions in most
cases. A possible interesting exception is pointed out in
\cite{Kharzeev1,Kharzeev2}. When relativistic ions undergo noncentral collision,
strong magnetic field is generated in the collision region. The
typical magnitude of this field is estimated as $\sqrt{eB} = 10 \div
100$ MeV, i.e. of the order of dynamical QCD scale. Correspondingly,
any studies of strongly interacting matter in heavy ion collisions
have to take the effects of this abelian magnetic field into
account. Of particular interest in this respect is the so called
Chiral Magnetic Effect (CME). The physics
behind it can be explained in several different but complementary
ways \cite{Kharzeev1}-\cite{Fukushima3}. Let us consider nonzero density of one flavor
of free massless quarks in external magnetic field. Suppose there
are unequal chemical potentials for left and right handed quarks:
$\mu_L \neq \mu_R$. When it can be shown
that a nonzero {\it classical} electric current flows along the
magnetic field (see \cite{Fukushima1} and references therein, see also \cite{sh} for another prospective):
 \be {\bf j} = \frac{e^2}{2\pi^2} \mu_5 {\bf B} \label{jB}
\ee where $2\mu_5 = \mu_R - \mu_L$.  The physical reason for this
chiral charge excess to electric charge current conversion is quark
magnetic moment interaction with the magnetic field (which is of
different sign for positively and negatively charged quarks)
together with the correlation of spin and momentum for chiral
fermions. Both sides of (\ref{jB}) have of course the same
transformation properties under P- and CP-parity conjugation. Many
different aspects of CME have been extensively discussed in the
literature and there is no doubt that CME is a robust theoretical effect.
However it is not a simple task to apply this clear physical picture to real processes described by nonperturbative QCD.
One of the most important questions on this way is about physical origin of chiral chemical potential $\mu_5$, which is absent in fundamental QCD Lagrangian. In original picture \cite{Kharzeev7} appearance of effective $\mu_5 \neq 0$ is a nonperturbative QCD effect, caused by interaction of quarks with
topologically nontrivial gluon field configurations {\it above} the phase transition. The physical explanation goes as follows. As is well known the topological charge in the QCD vacuum fluctuates as described by Veneziano-Witten formula \cite{v,w}
\be
\chi =\int d^4 x \lan G\tilde G(x) G\tilde G(0)\rrr \propto F^2 m_{\eta'}^2
\label{vw}
\ee
where the nonperturbative parameter in the r.h.s. scales as $\Lambda_{QCD}^4$ which means that topological charge fluctuates over Euclidean 4-volumes of typical size determined by nonperturbative QCD scale. It is worth stressing that these fluctuations are {\it quantum}, i.e. the states of different topological charge are to be summed over for whatever Euclidean 4-volume $V$ and one always has
\be
 \int_V d^4 x \lan G\tilde G(x) \rrr = 0
 \label{q1}
 \ee
 In other words, (\ref{q1}) vanishes because local average $\lan G\tilde G(x) \rrr = 0$ and not due to the presence of integration over the volume $V$.
 There is no special space-time fluctuation pattern in the problem other than the correlator (\ref{vw}) (and higher ones).

 The situation however may change at nonzero temperature/density. Since the Euclidean $\mathbb{O}(4)$ invariance of the vacuum is broken in this case, one can think of different fluctuation patterns in spatial and in temporal directions. Moreover, since in real collision experiments external conditions are time-dependent they can play a dual role of the background and of a measuring device. In other words the meaning of averaging in (\ref{q1}) changes: one has to integrate only over those field excitations which are present at a given Minkowski 3-volume for a given time period and the problem becomes essentially non-stationary in this sense. One can say that the average over fields $\lan .. \rrr$ becomes $V$-dependent. Such quantity - physically corresponding to a "single event" - can in principle be non-vanishing. Of course it is natural to expect that random character of fluctuations leads to zero result for (\ref{q1}) after averaging over many events.

The CME is often considered as a reasonable explanation of outgoing
particles electric charge asymmetry observed at Relativistic Heavy Ion Collider (RHIC)
\cite{Voloshin1} - \cite{Wang} in $\sqrt{s_{NN}} = 200$ GeV Au+Au and Cu+Cu collisions.
 The latter effect can be described as
follows. For noncentral collision one can fix the reaction plane by two vectors: beam momentum and impact parameter (without loss of generality this is always chosen as 12 plane in the present paper and no adjustment angle $\Psi_{RP}$ is introduced). Thus angular momentum of the beams (and the corresponding magnetic field) is oriented along the axis 3. The azimuthal angle $\phi \in [0,2\pi)$ is defined in the plane 23. With this notation, in any particular event one studies charged particles distribution in $\phi$ using the following conventional parametrization
\be
\frac{dN_{\pm}}{d\phi} \propto 1+2v_{1,\pm}\cos\phi + 2v_{2,\pm} \cos2\phi + 2a_{\pm} \sin\phi + ...
\label{ne}
\ee
 The coefficients $v_{1,\pm}$ and $v_{2,\pm}$ account for the so called directed and elliptic flow. They are believed to be universal for positively and negatively charged particles with good accuracy. The coefficients $a_+$ and $a_-$ describe charge flow along the third axis, i.e. normal to the reaction plane. This P-parity forbidden correlation between a polar vector (electric current) and the axial one (angular momentum) is considered as a signature of P-parity violation in a given event with $a_\pm \neq 0$. On the other hand, the random nature of the process dictates $\lan a_+\rrr_e = \lan a_-\rrr_e = 0$ (there the averaging over events is taken).

Trying to construct a theory of the phenomenon one has first to choose adequate language. Since at the end the heavy-ion collision is a scattering problem, the ultimate framework would be S-matrix and inelastic scattering amplitudes formalism with two colliding ions as incoming particles. Due to extreme complexity this way seems to be totally hopeless. Instead one uses some effective theories like hydrodynamics to predict distribution of outgoing particles. In the particular problem of charge fluctuations asymmetry the crucial point distinguishing different theoretical models is whether the currents of interest are treated as classical or as quantum. In the former case one makes use of the expression (\ref{jB}) as classical equation. The quantum nature of the problem here is hidden in a theory for $\mu_5$ and corresponding correlators and fluctuations for this effective chiral chemical potential. In the later case one is to consider quantum averages like $\lan \Omega | j_\mu |\Omega \rrr$, $\lan \Omega | j_\mu j_\nu |\Omega \rrr$ etc. and to understand (\ref{jB}) as operator relation.
 However if one takes diagonal matrix element of (\ref{jB}) in the vacuum the answer is of course trivial: $\lan 0 | {\bf j} | 0 \rrr=0$ even for nonzero external magnetic field. The absence of net electric current is directly related to the fact that fundamental QCD Lagrangian contains no such quantities as $\mu_L$ or $\mu_R$.

 We discuss three basic complementary ways to address quantum nature of CME in this paper:
  \begin{enumerate}
  \item
  To make use of P-parity violation probed by local order parameters
  \item
  To consider CME in quantum measurement theory framework
   \item
   To study $\mbox{P-odd} \times \mbox{P-odd}$ contributions to P-even observables.
   \end{enumerate}
We discuss all these approaches in the present paper and start with the first one in the next Section which is phenomenologically the simplest.

 \section{P-parity violation probed by local order parameters}

 As is well known quantum field theoretical averages of local operators have typically the following leading contribution:
\be
\lan \Omega | {\cal O}(x) |\Omega \rrr \propto c \cdot \Lambda^{d_{\cal O}}
\label{iu}
\ee
where $\Lambda$ is ultraviolet cutoff and numerical constant $c$ is generally non-vanishing if $c=0$ is not protected by some symmetry. Therefore the crucial step in the discussed problem is to model transition from local microscopic current $j_\mu$ to nonlocal macroscopic one $J_\mu$.
It is done by taking the matrix elements of the current $j_\mu$  over the medium degrees of freedom $|\Phi \rrr$ from full state vector $| \Omega \rrr = |\Phi \rrr \otimes |\phi \rrr$:
\be
j_\mu(x) = \bar\psi \gamma_\mu \psi(x) \leftrightarrow J_\mu \propto \lan \Phi | \> \int dx \rho_V(x) j_\mu (x) |\Phi \rrr
\ee
Here the function $\rho_V(x)$ defines the measure of integration over "physically infinitesimal volume", as is usual in condensed matter physics.

The second important ingredient is the existence of the medium itself. For phenomenological  purposes it is not important what particular
kind of microscopic description for the medium is chosen. What does matter is Lorentz symmetry breaking following from the existence of a distinguished frame - the medium rest frame. In the simplest cases of uniform medium characterized by nonzero temperature/density it is usually parameterized by a unit vector $u_\mu$ - the medium four-velocity, so that for applied uniform electromagnetic field one has the standard text-book answer for induced current
\be
\lan \phi |  J_\mu | \phi \rrr \propto u^\nu F_{\mu\nu}
\ee

We say about local parity violation in a state $|\Omega \rrr$ when a local parity-odd operator ${\cal O}(x) = - P {\cal O}(x) P^\dagger $ has nonzero expectation value
in this state
\be
\lan \Omega | {\cal O}(x) |\Omega \rrr \neq 0
 \ee
for example $\lan \psi^\dagger \gamma_5 \psi \rrr \neq 0$. The condition of locality here is important. Operationally it means that the operators and their products are defined at the scale $a\sim \Lambda^{-1}$ where $\Lambda$ is ultraviolet cutoff. For nonlocal averages, on the other hand, it is not a problem to have nonzero P-odd matrix element, e.g. $ \lan j_0(x) j_3(y) \rrr$. The medium, characterized by finite coherence length, brings physical meaning to this nonlocality. For example, in a medium with applied uniform electromagnetic field nothing forbids to have
P-odd correlation between axial current divergence and the vector current:
\be
\lan \phi | J_\mu \> \partial J^5 |\phi \rrr \propto u^\nu {\tilde F}_{\mu\nu}
 \label{j5j}
 \ee
 where ${\tilde F}_{\mu\nu} = \frac12 \epsilon_{\mu\nu\alpha\beta} F^{\alpha\beta}$.

To feel the physical meaning of (\ref{j5j}) let us imagine radial distribution of velocities ${\bf v}$ of the matter in a uniform magnetic field ${\bf B}$. If the divergence $\partial J^5$ is also uniform in the ("fireball") volume, the charge density is to be of different sign above and below the reaction plane:
  \be
\lan \phi | J_0  \> \partial J^5 |\phi \rrr \propto {\bf v} \cdot {\bf B}
 \label{j5j2}
 \ee
 In medium rest frame characterized by $u_\mu = (1,0,0,0)$ for uniform magnetic background, the electric current ${\bf J}$ flows along the magnetic field ${\bf B}$.

 It seems quite natural to interpret (\ref{j5j}) in the following way: as soon as the concept of a medium can be applied to the discussed problem
 one can easily construct classical nonzero local P-odd parameters without specifying any particular "chiral microscopy".
The medium (manifested by existence of the selected frame) is crucial in two aspects: first, it allows to consider meaningful local objects and not badly divergent quantities like (\ref{iu}) and second, by Lorentz invariance breaking it provides invariant meaning for the electric and magnetic fields, thus making possible correlations between local (in macroscopic sense!) operators of different parities.
We also see here the importance of the uniformity condition: if $\partial J^5$ is short-correlated, there is no net effect. This brings us back to the question about dynamical scales hierarchy. 

There is a deep question behind the above consideration: if the (microscopic) current non-conservation is anomalous (e.g. in (\ref{j5j})) - how is this fact encoded in equations for macroscopic, effective currents? We leave aside the discussion of this important point and refer an interested reader to \cite{son} where this question is addressed in hydrodynamic setup.

 From heavy ions collision point of view the P-parity violating average (\ref{j5j}) is not an observable by itself. The reason is physically clear: instead of measuring components of vector (electric) current and axial (chiral) current and studying their correlation, only the quantities of the former kind are being measured (in the form of final particles electric charge distribution). As for the latter quantities related to chiral charge - it is assumed
 that the quark-gluon medium created after collision of two heavy ions plays itself a role of the measuring device
performing effective measurement of topological charge in the corresponding space-time region.
It should be mentioned that this is
a rather strong assumption: to say that one part of some quantum system can measure (in classical sense) the state of another part of the same system
means in fact to address some scenario for decoherence of the subsystems and information loss. To model this effect one has to adopt the language of quantum measurement theory. This is done in the next Section.

 \section{CME in quantum measurement theory framework}

 It is possible to understand (\ref{jB}) as a correlation between preferred direction of outgoing electric charge distribution asymmetry and the magnetic field in a particular event. The sign of this P-parity odd asymmetry is fixed by the sign of effective $\mu_5$ in this event (and of course varies randomly from event to event due to topological neutrality of QCD vacuum). The quantitative theory would require information about distribution function of effective $\mu_5$.

A simple quantum-mechanical analogy can be useful to illustrate the point. In one-dimensional bound state problem with P-parity even potential $V(x)=V(-x)$ one has $\lan x \rrr = \int x dx |\psi_0(x,t)|^2 =0$ where $\psi_0(x,t)$ is the ground state P-parity even wave function. On the other hand, performing particle position measurements on ensemble of $N$ identical systems all in the ground state one gets sequence of positive and negative numbers $x_1, x_2,..,x_N$ (with some uncertainties determined by the measuring device properties). Quantum mechanics does not predict the result of a single measurement, but guarantees $\lan x \rrr = \lim\limits_{N\to\infty} \frac{1}{N} \sum_{i=1}^N x_i = 0$. For each measurement with the outcome $x_i \neq 0$ one can say that P-invariance is broken in this particular experiment, "event-by-event". In this simple case "breaking" is clearly of statistical origin and has nothing to do with dynamics - i.e. properties of the potential $V(x)$. Therefore it is common in quantum mechanics not to use such terminology and compute instead nonzero P-parity even observables, such as, e.g. $\lan x^2 \rrr = \int x^2 dx |\psi_0(x,t)|^2  \neq 0$, characterizing the pattern of quantum fluctuations. What is however important is the textbook average over events / average over probability density equivalence.

 By way of another simple analogy consider a system of massless quantum fields subject to boundary conditions at typical distance scale $L$ characterized by a unit 3-vector ${\bf n}$. To be concrete one can think of electromagnetic Casimir vacuum between parallel plates at distance $L$ with ${\bf n}$ being normal to the planes. Let this vector smoothly fluctuates in random directions with typical frequency $\omega$, which is assumed to be much smaller than $c/L$. One studies the quantum average of energy-momentum tensor for the fields, $\lan T_{\mu\nu}(x)\rrr $. Since the problem is quasi-stationary, the general answer is given by \be \lan T_{\mu\nu}(x)\rrr = a(x)\> g_{\mu\nu} + b(x)\> n_\mu n_\nu + {\cal O}(\omega L /c)
 \label{tt}
 \ee
  On the other hand, performing an average over time period $T \gg \omega^{-1}$ one should have \be \frac{1}{T}\int_0^T dt \>\lan T_{\mu\nu}(x)\rrr = \bar a(x)\> g_{\mu\nu} \label{tt2}\ee
  since no memory has remained about the particular direction the vector ${\bf n}$ is pointing to. Thus experiments with the detector time resolution $\omega T \ll 1$
  will observe $\mathbb{O}(3)$ violating local answer (\ref{tt}) while those over long time scales $\omega T \gg 1$ will see $\mathbb{O}(3)$ respecting answer (\ref{tt2}).
  It is of crucial importance that some physical process with the typical life time scale comparable or larger than the plasma life time does exist and it is responsible for creation of P- and CP-odd domains in the dense and hot matter in Minkowski space-time. It seems to be rather subtle point in this case how a relation between Euclidean expression (\ref{vw}) and Minkowskian dynamics should look like. In any case the existence of scale separation
  between the process dynamical scales and the measuring device ones is necessary for the whole picture to make sense.

Since detailed picture of the discussed microscopic quantum/classical interplay is beyond us, our attitude here is purely phenomenological.
We define the effective $\eta$-dependent current $J_\mu(x,\eta)$ as
\be
J_\mu(x,\eta) = \lan \Omega_{\eta} | j_\mu(x) |\Omega_{\eta} \rrr
\label{jjj}
\ee
where electric current $j_{\mu}(x) =   \bar{\psi}(x) Q \gamma_\mu \psi(x)$ with quarks charge matrix $Q = \mbox{diag}(2/3, -1/3, -1/3)$.
The state $ |\Omega_{\eta} \rrr$ is characterized by
\be
\lan \Omega_{\eta} |\int_V d^4 y \> \partial j^5(y) |\Omega_{\eta} \rrr = \eta
\ee
It is physically obvious that $J_\mu(x,\eta)$ must be an odd function in $\eta$ and
\be
\int_{-\infty}^{\infty} d\eta J_\mu(x,\eta) = 0
\ee
Since by assumption each event is characterized by some value of $\eta$, positive or negative with equal probability, this corresponds to "averaging to zero" over many events.

To proceed further it is convenient to use the formalism of partial partition functions:
 \be
Z = \int D\Phi \exp(-S[\Phi]) \prod\limits_i \int d \eta_i \> \tilde\delta(\eta_i - O_i[\Phi])
\ee
where $S[\Phi]$ is the standard Euclidean QCD action, $\Phi$ stays for dynamical quark and gluon fields $A,\bar\psi, \psi$ and $O_i[\Phi]$ is a gauge-invariant operator made of these fields.
We approximate the real detector with finite resolution by the choice of the "detector function" $\tilde\delta(x)$ in Gaussian form:
 \be
 \tilde\delta(\eta) = \frac{1}{2\pi} \int_{-\infty}^{\infty} d\lambda \exp(-\lambda^2 l^2/2 + i\lambda \eta )
\ee
so that $\int_{-\infty}^{\infty} d \eta  \tilde\delta(\eta) = 1$.

We are interested in a value of the electric current (\ref{jjj}). For exactly conserved axial current $\partial j^5 = 0$ one would have
$
\lan \Omega | j_\mu(x) \cdot\partial j^5(y) |\Omega \rrr = 0
 $.
Due to (electromagnetic) anomaly however the result reads (for isovector components)
$$
i \int dx \> e^{iq(x-y)} \> \lan \Omega | j_\mu(x) \cdot\partial j^{5,a} (y) |\Omega \rrr =
$$
\be
= \T [Q^2 t^a] \cdot \left( - \frac{N_c}{4\pi^2} \right) \cdot q_\nu {\tilde F}_{\mu\nu}
 \label{wei}
 \ee
where $t^a$ are generators of flavour $SU(2)$ or $SU(3)$.

For singlet current the anomaly gets gluon contribution
\be
\partial j^5 = -{\mbox{Tr}} [Q^2] \> \frac{N_c}{4\pi^2} F_{\mu\nu}{\tilde F}^{\mu\nu}
- \frac{N_f}{16\pi^2} \T G_{\mu\nu} {\tilde G}^{\mu\nu}
\ee
(notice that for uniform magnetic field $F_{\mu\nu}{\tilde F}^{\mu\nu} =0$) and computing
\be
J_{\mu}(\eta, x) = \frac{1}{Z} \int D \Phi \> j_\mu(x) \> \tilde\delta\left(\eta - n_V \right) \>\exp(-S[\Phi])
\label{cur}
\ee
where \be
n_V =  \int_V d^4 y \partial j^5 = - \frac{N_f}{16\pi^2} \int_V d^4 y \T G_{\mu\nu}{\tilde G}^{\mu\nu}\ee
 at the leading order of
the cluster expansion
\be
 \lan A\exp B \rrr
 \approx \lan AB\rrr \exp(\lan B^2 \rrr/2) \ee
  valid for $\lan A \rrr = 0 $ and $\lan B \rrr = 0$,
one gets in this approximation
\begin{widetext}
\be
J_{\mu}(x,\eta) = -{\mbox{Tr}} [Q^2] \> \frac{N_c}{4\pi^2}\> \frac{\eta e^{-\eta^2/2L^2}}{\sqrt{2 \pi L^6} }  \cdot \left[\int \frac{d^4 q}{(2\pi)^4} e^{iqx} f_V(q) iq^\nu \right] \cdot {\tilde F}_{\mu\nu}
\label{fincur}
\ee
\end{widetext}
Here $L^2 = l^2 + \lan  n_V^2 \rrr$ and the formfactor is given by $f_V(q) = \int_V d^4 y \exp(-iqy)$. In the infinite volume limit $\chi = \lim\limits_{V\to \infty} \lan  n_V^2 \rrr /V N_f^2 $ where $\chi$ is the standard topological susceptibility.

The expression (\ref{fincur}) deserves a few comments. First, the right hand side of (\ref{fincur}) is odd function of $\eta$ as it should be, and  at small $\eta$ the current is linear in $\eta$. If the point $x$ is far from $y\in V$ the current vanishes due to formfactor $f_V(q)$, i.e. the current flows only in the interaction volume $V$.
On the other hand, if $x\in V$ and $V$ is large enough to neglect surface terms, the current also vanishes as it should be for any finite-volume effect. The volume scaling $\lan  n_V^2 \rrr \sim V$ for the phase with finite correlation length is another manifestation of the same fact.

It is worth mentioning that the maximal current is reached at $\eta \sim L$ and decrease as $J^{max} \propto B/\tau L^2$ (where $\tau \sim V^{1/4}$).
This result seems counter-intuitive. Indeed, a naive picture would suggest that stronger fluctuations of topological charge $\lan  n_V^2 \rrr$ are to correspond to stronger currents $J_{\mu}(x,\eta) $. This in fact is not the case. Rough physical explanation follows from (\ref{wei}): since the product of $j_\mu$ and $\partial j^5$ is fixed by electromagnetic anomaly (i.e. by the magnitude of external abelian field $F_{\mu\nu}$) large  $\partial j^5$ corresponds to small $j_\mu$ and vice versa. Let us remind that according to the lattice data \cite{ddl} the magnitude of topological charge fluctuations experience rather sharp drop above the deconfinement transition. According to the above it means the effective {\it enhancement} of maximal possible electric current fluctuations!
Of course at too small $\lan  n_V^2 \rrr $ Gaussian approximation (neglect of higher order correlators) we have used is to break down.

It is seen that the discussed effect is a result of subtle interplay  between strong and electromagnetic anomalies (see related remarks in \cite{Fukushima1}). While the later one is responsible for correlation between vector and axial currents, the former anomaly provides non-conservation of axial charge due to topological nonperturbative gluon fluctuations. The question about $\mu_5$ distribution addressed in the introduction is translated here into the question about $\eta$ distribution for experimental events.

\section{Charge fluctuations asymmetry and polarization operator}

 Perhaps the most logically consistent way is to study transition matrix elements of (\ref{jB}) between states of opposite P-parity.
 This corresponds to:
\be
\lan \Omega| j_i \> j_k |\Omega \rrr  \to \sum_A \lan \Omega | j_i |A \rrr \lan A |j_k |\Omega \rrr
\label{jsq2} \ee
where the states $|\Omega\rrr$ and $|A\rrr$ have opposite P-parities and $\lan A | \Omega \rrr = 0$.
Of course the expression (\ref{jsq2}) is nothing but the electromagnetic polarization operator in the state $|\Omega\rrr$ saturated by particular states in spectral expansion.

This line of reasoning has been addressed in the literature before.
Local averages like $\lan j_\mu^2(x) \rrr$ were computed in pioneering studies of CME on the lattice \cite{Buividovich1,Buividovich2} and many interesting patterns were found. Later nonlocal averages $\lan j_\mu(x) j_\nu(y) \rrr$ are computed \cite{Buividovich3,Buividovich4}. We find it worth reminding once again that since the typical correlators we are interested in are given by dimension six operators, their local matrix elements are strongly UV-singular
\be
\lan j_\mu^2(x) \rrr_F  \propto \Lambda^6 + F^2 \Lambda^2 + \mbox{UV-finite}
\label{uv}
\ee
where $\Lambda$ is UV-cutoff and $F$-external field strength. Even subtracted average $\lan j_\mu^2(x) \rrr_F - \lan j_\mu^2(x) \rrr_0$ is divergent. This problem is overcome in numerical lattice calculations, but present analytical challenge for any attempt to describe CME in terms of local matrix elements. To our view this is a clear signal about intrinsic nonlocal nature of the discussed phenomenon.

Polarization operator in the CME context is studied in \cite{Fukushima2}. There are two main differences between our approach and that of the cited paper.
First the regular contribution (given by polarization operator in magnetic field) and CME-contribution (proportional to $\mu_5$) are separated from the beginning in \cite{Fukushima2} (in some sense, quantum currents are superimposed on top of the classical current (1)). We follow another logic and consider polarization operator as the only source of asymmetric charge fluctuations, but extract a particular formfactor from it, which corresponds to negative parity intermediate states. Second, the expression for charge fluctuations observable as a functional depending on polarization operator is different in our paper from that of \cite{Fukushima2}. We will make more comments on that below.

In this section we discuss $\mbox{P-odd} \times \mbox{P-odd}$ contributions to P-even observable, the role of which is played by current correlator $\lan j_\mu j_\nu \rrr$.
It seems physically clear that  this object should contain some information about charge distribution (\ref{ne}). The exact form of this correspondence is however far from trivial. One could think of several ways to relate these quantities. Before presenting our approach to this problem let us mention other methods used in the literature. First, we notice that the current in $\phi$-direction is given by
\be
{\bf e}_z j_3 + {\bf e}_y j_2 = \sqrt{j_3^2 + j_2^2}({\bf e}_z \sin\phi + {\bf e}_y \cos\phi)
\ee
and the corresponding charge difference from (\ref{ne}) is
\be
\left\lan \int \frac{d(N_+ - N_-)}{d\phi} \> d\phi \>\int \frac{d({N'}_+ - {N'}_-)}{d\phi'}\> d\phi' \right\rrr_e
\label{nn}
\ee
where by the brackets $\lan ... \rrr_e$ we denote the average over events. One has
$
\lan (a_+ - a_-)^2 \rrr_e \propto \lan j_3^2 + j_2^2 \rrr
$
where the current product is assumed to be local. This is very close (but different) to the definition used in \cite{Buividovich1}.
It is natural to expect that positive definite $
\lan (a_+ - a_-)^2 \rrr_e $ should be nonzero even without any magnetic field.

Another relation is suggested in \cite{Fukushima2}. It is written in terms of event average of the cosine, where $\alpha, \beta = +, -$ and $N_\pm$ is the total number of outgoing particles of a given charge:
\be
\lan \cos(\phi_\alpha + {\phi_\beta}') \rrr_e \propto \frac{\alpha\beta}{N_\alpha N_\beta} (j_2^2 - j_3^2)
\ee
where, up to some background terms
\be
\lan \cos(\phi_\alpha + {\phi_\beta}') \rrr_e = \lan v_{1,\alpha} v_{1,\beta} \rrr_e - \lan a_\alpha a_\beta \rrr_e
\label{p9}
\ee
Assuming charge independence of $v_{1,\alpha} $ and equal numbers of particle species $N_+ = N_- = N$ one gets
 $\lan (a_+ - a_-)^2 \rrr_e \propto \lan j_3^2 - j_2^2 \rrr$ if one neglects $v_{1,\alpha}$ term with respect to $a_\alpha$ term. In fact, the leading term, which is always contained in $j_3$ component, coincides for both expressions, while the procedure of taking into account fluctuations in the reaction plane is different.

 In this paper we use alternative signature provided by charge density fluctuations and not spatial components of the currents.
 An attractive feature of this quantity is that it is well defined even in the static limit.
To this end
consider electric charge in some spatial volume $V$ at temperature
$T$: \be eQ_{V} = e\int_V d{\bf x}  \> j_0(x) \ee Since we work in zero density approximation the quantum average
of this object vanish: \be \lan Q_{V} \rrr = 0 \ee This is not the case for
its square: \be \lan Q_{V}^2 \rrr = - {\hat \kappa} \int_V d{\bf x} \> \int_V d{\bf x'} \>
\Pi_{44}(x, x') \label{qq} \ee
where $\Pi_{44}(x, x')$ is Euclidean polarization operator in constant external field $F_{\mu\nu}$ and at temperature $T$ Wick-rotated from the standard Minkowski expression $\Pi^{(M)}_{00}(x,x')$:
\be \Pi^{(M)}_{\mu\nu}(x,x') = i \lan T\{j_\mu(x)j_\nu(x')\}
\rrr_{F,T} \ee with $j_\mu = \bar\psi Q \gamma_\mu \psi$; $\Pi^{(M)}_{\mu\nu} \leftrightarrow \Pi^{(E)}_{\mu\nu}$,
notice the sign convention (\ref{qq}) corresponding to positive-definite $\lan Q_{V}^2 \rrr $ in the static limit.
In the standard way we denote
 \be \Pi_{\mu\nu}(q) = \int d^4 x \> e^{-iq(x-x')} \>
\Pi_{\mu\nu}(x,x') \label{po} \ee
with $\mu,\nu =1,2,3,4$ and ${\bf q} = (q_1,q_2,q_3)$, $q_\bot = (q_1,q_2)$.

The operator ${\hat \kappa}$ in (\ref{qq}) accounts for temporal profile of the process.
In terms of momentum space components, (\ref{qq}) takes the following form
\be
 \lan Q_{V}^2 \rrr = - \int \frac{dq_4}{2\pi} \kappa(q_4) \int \frac{d{\bf q}}{(2\pi)^3} |F_V({\bf q})|^2 \Pi_{44}({\bf q}, q_4)
 \label{polatas}
\ee
where the form-factor $F_V({\bf q}) = \int_V d{\bf x} \exp(i{\bf qx})$ keeps information about spatial profile of
the volume $V$, while the temporal factor $\kappa(q_4) = \int d\tau g(\tau) \exp(iq_4\tau)$ encodes temporal (in Euclidean sense)
profile. For finite temperature case we consider here the standard Matsubara replacements $q_4 \to \omega_n = 2\pi n T$ and $(2\pi)^{-1}\int dq_4 \to T\sum_n$ are to be performed. The choice $g(\tau) = T$ we will adopt in the rest of the paper physically corresponds to the static limit where only the lowest Matsubara frequency $n=0$ contributes:
\be
 \lan Q_{V}^2 \rrr_{st} = - T  \int \frac{d{\bf q}}{(2\pi)^3} |F_V({\bf q})|^2 \Pi_{44}({\bf q}, 0)
 \label{polat}
\ee
Using the expressions from Appendix it can be checked that in thermodynamic limit $V\to \infty$ without external field
one reproduces standard Stefan-Boltzmann answer for elementary fermions  $\lim\limits_{V\to \infty} \lan e^2 Q_{V}^2 \rrr_{st} /V = e^2 T^3 /3$.
In case of quarks one should of course understand $eB$ as $q_f eB$ and introduce additional trace over flavors with the factor $N_c Q^2$:
$
\Pi_{44}^{eB, T} \to N_c \sum_f q_f^2 \Pi_{44}^{q_f eB, T}
$. For the sake of brevity we will use the simple notation as for elementary fermions of unit electric charge having in mind the necessity to make the replacement discussed above in the final answers.

In the limiting case of no background $B=0, T=0$ one has $\Pi_{44}({\bf q}, q_4) = {\bf q}^2 \Pi({q}^2)$ and, at the leading order, for large 4-volumes  $V_4$:
\be
\lan Q_{V}^2 \rrr_{B=0,T=0} \propto {\Pi}'(0)\cdot V_4^{-1/2}
\ee
where the condition of gauge invariance $\Pi(0)=0$ has been taken into account and the volume $V_4 = R^3\times t$ is assumed to be uniform: $R\sim t$.
Thus the expression (\ref{qq}) is UV-safe and vacuum charge fluctuations in a given space-time region is purely finite-volume effect.

We can now come back to the definition (\ref{qq}) and rewrite the
coordinate integration in cylinder coordinates with the axis 1 as
the polar axis and angle $\phi$ defined in the 23 plane. This is the same notation as in (\ref{ne}), notice that in the standard setup
azimuthal angle is usually defined in the plane $12$. This allows
to represent the form-factor $F_V({\bf q})$ as \be F_{V}({\bf q}) =
\int dx_1\> e^{iq_1x_1} \int_0 \rho d\rho \int_0^{2\pi} d\phi \>
e^{i{\bar q}\rho} \label{fot} \ee where ${\bar q}\rho = q_2 x_2 +
q_3 x_3 = q_2 \rho \cos\phi + q_3 \rho \sin\phi$ and the structure
of integration upper limit is determined by the chosen model for
spatial distribution (sharp boundary, smoothed boundary, Gaussian
shape, exponential shape etc). The $\sin \phi$ - mode in Fourier
expansion of (\ref{fot}) is multiplied by the following coefficient
\be c_1 = (1/\pi)\int_0^{2\pi} d\phi \sin\phi e^{i{\bar q} \rho} =
\frac{2i q_3}{\hat q}\> J_1({\hat q}\rho) \ee where ${\hat q} =
\sqrt{q_2^2 + q_3^2}$. Thus we have for expansion of (\ref{polat}) in harmonics: \be \lan
Q_{V}^2 \rrr = ... + \int_0^{2\pi} d\phi \sin\phi \int_0^{2\pi}
d\phi' \sin\phi' \lan (q^a_{V})^2 \rrr + ... \label{pol} \ee where $\lan
(q^a_{V})^2 \rrr$ is given by the same expression (\ref{polat}) with the
change $F_V({\bf q}) \to f_V(q_1,q_2,q_3)$ where \be f_V(q_1,q_2,q_3) =  \frac{2i q_3}{\hat
q}\>\int dx_1\> e^{iq_1x_1} \int_0 \rho J_1({\hat q}\rho) d\rho \ee
In the same way $\lan (q^{v_1}_{V})^2 \rrr$ corresponds to the exchange $q_3 \leftrightarrow q_2$ and $\sin \phi \leftrightarrow \cos \phi$.
Making use of (\ref{ne}), (\ref{nn}) and (\ref{pol}) we obtain the following relation
for the asymmetry \be
\lan q_{V}^2 \rrr = \lan (q^a_{V})^2 \rrr - \lan (q^{v_1}_{V})^2 \rrr = -\sum\limits_{\alpha,\beta = \pm} \alpha \beta \cos(\phi_\alpha + {\phi_\beta}')
\ee
\begin{widetext}
\be
\lan
q_{V}^2 \rrr =
N^2\cdot \left( \lan (a_+
- a_-)^2 \rrr_e - \lan (v_{1,+}
- v_{1,-})^2 \rrr_e \right) = T\int
 \frac{d{\bf q}}{(2\pi)^3} \frac{q_3^2 - q_2^2}{q_3^2 + q_2^2} \left| \int dx_1\> e^{iq_1x_1} \int_0 \rho J_1({\hat q}\rho) d\rho\right|^2 \> \Pi_{44}({\bf q}, 0)
 \label{as}
\ee
\end{widetext}
It is obvious that the above expression has to be proportional to magnetic
field since there is no other $O(3)$-violating factors in the
problem. The effect we are looking for corresponds to {\it
strong enhancement} of (\ref{as}) in external magnetic field and hence,
from experimental point of view, strong dependence of (\ref{as}) on centrality. It is to be stressed that the multiplicity factor $N^{2}$
is by itself strongly centrality-dependent. This dependence is kinematical and has nothing to do with magnetic field dependence of $\lan q_V^2 \rrr$.
Only the latter lies at the heart of CME.

\section{General structure of polarization operator}

In this section we analyze general structure of polarization
operator in the background of nonzero temperature and magnetic
field. As is clear from the above discussion, this is a necessary
prerequisite before one can compute the charge fluctuation
asymmetry (\ref{as}).

First of all let us make a few general comments about space-time dependence of current-current correlator.
In confinement phase (i.e. at sufficiently low temperatures) at large distances and for weak magnetic field one expects general structure of Euclidean polarization operator of the following form
\be
\lan j(x) j(x') \rrr \propto e^{-m_\rho |x-x'|} + C(B) \cdot e^{-m_\pi |x-x'|}
\ee
with $C(B) \propto B^2$. This interesting effect of different parity states mixing in external field is similar to the one observed long time ago in \cite{ioffe}
at finite temperature. The long-distance correlations are thus saturated by the lightest degrees of freedom (i.e. pions in the confinement phase).
On the other hand, in deconfinement phase at strong fields, if Larmor radius is much smaller than $\Lambda_{QCD}^{-1}$ no quarks can propagate in transverse direction at all:
\be
\lan j(x) j(x') \rrr \propto e^{-eB(x-x')^2_\bot /2}
\ee
Large-$N_c$ suppressed transverse correlations are possible only due to gluon degrees of freedom.

We
confine our attention in what follows to a particular case of purely
magnetic constant abelian background field $F_{\mu\nu}$ in the thermal bath rest frame at nonzero temperature $T$.
  We have chosen
$F_{12} = -F_{21} = B$, i.e. magnetic field is directed along the
third axis. The temperature effects break Lorentz-invariance and the
physical answers depend on 4-vector $u_\mu$ which
represents four-velocity of the thermal bath. It is normalized as
$u_\mu u^\mu = 1$. In the present paper we take zero chemical
potential $\mu = 0$. It is to be noticed that many general
conclusions concerning the structure of polarization operator stay
intact for $\mu\neq 0$ since the latter is associated with the same
four-vector $u_\mu$ given by $u_\mu = (1,0,0,0)$ in the medium rest
frame.

 The polarization operator (\ref{po}) is a rank two tensor
depending on two polar vectors $q_\mu$ and $u_\mu$ and antisymmetric
tensor $F_{\mu\nu}$. The general decomposition of (\ref{po}) in
terms of independent tensors was extensively studied in the literature starting from \cite{shabad1,bk}, see \cite{shabad2} for recent exposition and
\cite{kb} for a useful collection of references.
Generally, one is to deal with
$4\times 4 = 16$ independent tensor structures, built by multiplying
the four independent base vectors $ q_\mu, u_\mu, q^\alpha
F_{\alpha\mu},q^\alpha F_{\alpha}^{\beta}F_{\beta\mu}$. It can be
shown however that general requirements of being transversal \be
q^\mu \Pi_{\mu\nu}(q) = q^\nu \Pi_{\mu\nu}(q) = 0 \ee and Bose
symmetric $\Pi_{\mu\nu}(q) = \Pi_{\nu\mu}(-q) $ together with
generalized Furry's theorem \cite{shabad1} \be \Pi_{\mu\nu}(q,u,F) =
\Pi_{\mu\nu}(q,-u,-F) \ee reduce the number of independent tensor
structures to six. Two of them are field-independent, the other two
depend on $F_{\mu\nu}$ linearly and the last two - quadratically
(notice that our numeration of the tensors is different from the one
adopted in \cite{shabad1}). Their explicit form reads
$$ \Psi^{(1)}_{\mu\nu} = q^2 \delta_{\mu\nu} - q_{\mu}q_{\nu}$$
$$ \Psi^{(2)}_{\mu\nu} = (q^2 u_\mu - q_\mu (uq))(q^2 u_\nu - q_\nu (uq))
$$
$$ \Psi^{(3)}_{\mu\nu} = (uq)(q_\mu F_{\nu\rho}q^\rho - q_\nu
F_{\mu\rho}q^\rho + q^2 F_{\mu\nu})
$$
$$ \Psi^{(4)}_{\mu\nu} = (u_\mu F_{\nu\rho}q^\rho - u_\nu
F_{\mu\rho}q^\rho + (uq) F_{\mu\nu})
$$
$$ \Psi^{(5)}_{\mu\nu} = F_{\mu\rho}q^\rho F_{\nu\sigma}q^\sigma
$$
\be \Psi^{(6)}_{\mu\nu} = (q^2 \delta_{\mu\rho} - q_{\mu}
q_{\rho})F^\rho_\alpha F^{\alpha\sigma}(q^2 \delta_{\sigma\nu} -
q_{\sigma} q_{\nu}) \label{tensors} \ee The coefficient functions of
the decomposition \be \Pi_{\mu\nu}(q,u,F) = \sum\limits_{i=1}^6
\pi^{(i)}\cdot \Psi^{(i)}_{\mu\nu} \label{expansion}\ee depend on
$q^2$, mixed invariants $(uq)^2,$ $(qF)^2,$ $(uF)^2,$ $(qFu)^2$,
pure field invariants $F^2, F\tilde F$ and also the temperature $T$
and particle data, encoded in matrices $Q$ and $M$. The expression
(\ref{expansion}) allows to discuss current correlations asymmetries
in invariant way in any theory where the expression for polarization
operator can be obtained.

Having these general prerequisites let us come back to analysis of
correlation patterns. For our choice $F_{12} = B$ the invariants
$(uF)^2,$ $(qFu)^2$ and $ F\tilde F$ equal to zero. In what follows
we will be especially interested in a particular type of
contribution to $\Pi_{\mu\nu}(q) $ proportional to the tensor
structure $\Psi^{(7)}_{\mu\nu} $ given by the product of two axial
vectors \be
 \Psi^{(7)}_{\mu\nu} = \tilde
F_{\mu\rho}q^\rho \tilde F_{\nu\sigma}q^\sigma \label{psi7}\ee
It is not independent and one easily checks that
$\Psi^{(7)}_{\mu\nu} $ can be expressed as a linear combination of
(\ref{tensors}): \be
 q^2 \Psi^{(7)}_{\mu\nu} = \left(q^2 F^2 /2 - (qF)^2 \right) \Psi^{(1)}_{\mu\nu}
 +q^2 \Psi^{(5)}_{\mu\nu} + \Psi^{(6)}_{\mu\nu} \label{7} \ee

Let us consider tensor structure of the polarization operator in
more details. First of all, since we are interested only in diagonal 11, 22, 33, 44 and also 34 components in this paper, we have no contributions from $\pi^{(3)}$ and
$\pi^{(4)}$ because the tensors $\Psi^{(3)}_{\mu\nu}$ and
$\Psi^{(4)}_{\mu\nu}$ are antisymmetric and also vanish for $\mu=3, \nu=4$ in the chosen background field. Second, we notice that for
$\mu,\nu$ equal to 3 or 4, one has identically $\Psi^{(5)}_{\mu\nu} = 0$.
Adopting conventional notation: $q_\bot = (q_1,q_2)$, $q_{||} = (q_3,q_4)$
we can rewrite (\ref{expansion}) using (\ref{7}) as
\be
\Pi_{||}(q)= \pi^{(Q)}\cdot \Psi^{(1)}_{||} + \pi^{(T)}\cdot \Psi^{(2)}_{||}+ {\tilde\pi}^{(F)}\cdot \Psi^{(7)}_{||} \label{expansion2}\ee
where the new invariant functions are given by
$$
 \pi^{(Q)} = \pi^{(1)} - \left(q^2 F^2 /2 - (qF)^2 \right)\pi^{(6)}
 $$
 \be
 \pi^{(T)} = \pi^{(2)} \; ; \; {\tilde\pi}^{(F)} = q^2 \pi^{(6)}
 \label{pi1}
\ee
As for the diagonal correlators in 12-plane, one has
\be
\Pi_{\bot}(q)= \pi^{(Q)}\cdot \Psi^{(1)}_{\bot} + \pi^{(T)}\cdot \Psi^{(2)}_{\bot}+ \pi^{(F)}\cdot \Psi^{(5)}_{\bot} \label{expansion3}\ee
where $ \pi^{(Q)} $ and $\pi^{(T)} $ are defined by the same expressions (\ref{pi1}) while $\pi^{(F)}$ form-factor reads
\be
\pi^{(F)} = \pi^{(5)} - q^2 \pi^{(6)}
\label{pif}
\ee

 It is seen that the correlators of our interest can be decomposed into just three independent structures. The first, $\pi^{(Q)}$ corresponds to purely quantum fluctuations. It has nonzero limit at both $B\to 0$ and $T\to 0$, which coincides in this case with the textbook expression for polarization operator. The second structure, $\pi^{(T)}$ is responsible for thermal fluctuations. It vanishes at $T\to 0$. It is worth mentioning that both functions $\pi^{(Q)}$ and $\pi^{(T)}$ depend on temperature and external field (since the pattern of both quantum and thermal fluctuations is sensitive to the external conditions) and our notation corresponds rather to the limiting form of these functions.

We notice that the terms proportional to  $\pi^{(Q)}$ and  $\pi^{(T)}$  are identical in (\ref{expansion2}) and (\ref{expansion3}) up to obvious change of notation $|| \leftrightarrows \bot$. This is to be expected since quantum and thermal fluctuation are $\mathbb{O}(3)$-isotropic.
The only non-isotropic term (and the most interesting for us here) is the last terms: ${\tilde\pi}^{(F)}$ in (\ref{expansion2}) and ${\pi}^{(F)}$ in (\ref{expansion3}). The former one takes into account charge (and also the current component $j_3$) fluctuations induced by the external magnetic field. P-parity structure of this term is given by
$$
\begin{array}{cccccccc}
\delta_B \lan j_3 j_3 \rrr &=& {\tilde\pi}^{(F)} &\times &{\tilde F}_{3\rho}p^\rho &\times &  {\tilde
F}_{3\sigma}p^\sigma \\
\mbox{P-even}& = & \mbox{P-even}&\times & \mbox{axial}& \times & \mbox{axial}\\
\label{str}
\end{array}
$$
It is to be compared with the thermal contribution proportional to  $\Psi^{(2)}_{||}$:
$$
\begin{array}{cccccccc}
\delta_T \lan j_3 j_3 \rrr &=& \pi^{(T)}&\times & p_3 (up)&\times &  p_3 (up)\\
\mbox{P-even}& = & \mbox{P-even}&\times & \mbox{vector}& \times & \mbox{vector}\\
\end{array}
$$
This directly corresponds to our discussion in the introduction: in
the latter case the thermal fluctuations are distributed
isotropically in the thermal bath rest frame, while in the former
one there are electric currents fluctuating along the magnetic field.
The magnitude of these fluctuations is measured by the function
${\tilde\pi}^{(F)}$, and no physical principle forces it to vanish either
below or above critical temperature. Physically
${\tilde\pi}^{(F)}$ corresponds to P-odd
intermediate states in the polarization operator.

The function  ${\pi}^{(F)}$ entering (\ref{expansion3}) is a sum of two terms according to (\ref{pif}). This also is to be expected.
Charged particles flowing in the plane perpendicular to the magnetic field are deflected by the Lorentz force, and this {\it diamagnetic}
effect is taken into account by the form-factor  ${\pi}^{(5)}$. It is absent in $\Pi_{||}$. But particle's spin interacts with the field by means of ${\bf \sigma_{\alpha\beta} F^{\alpha\beta}}$ term in $\Pi_{||}$ as well as in $\Pi_{\bot}$ which results in the factor  $q^2 {\pi}^{(6)}$ in both expressions (\ref{expansion2}) and (\ref{expansion3}).
It is worth noting that according to our general logic the electric charge asymmetry is computed for the full expression for $\Pi_{44}$, not just from some part of it, proportional to   ${\tilde\pi}^{(F)}$. Thus it is legitimate to speak about CME-interpretation of the answer (\ref{as}) only in the limiting case when  ${\tilde\pi}^{(F)}$ provides dominant contribution. We discuss that in more details below.

\section{Model examples}

We analyze in this section two limiting cases where one can construct ${\tilde\pi}^{(F)}$ in explicit way. The first one corresponds to
weak magnetic fields in the confinement phase. In this case the intermediate states are hadron resonances of negative P-parity (see closely related discussion in \cite{mul}). However to select explicitely physical states making dominant contribution is far from trivial and the answer strongly depends on kinematics. We confine ourselves in this paper to the simplest case  keeping only three neutral $0^{-+}$ intermediate states: $\pi^0, \eta, \eta'$. Technically it is more convenient to consider from the very beginning matrix elements of vector currents between vacuum and these states in external field. Making use of the definition of off-shell vector-vector-axial form-factor ${\cal F}_\pi \equiv {\cal F}_{\pi^{0*}\gamma^*\gamma^*}(q^2,q_1^2,q_2^2)$ (see, e.g. \cite{nyf}) with $q=q_1+q_2$
$$
\int dx \int dy e^{iq_1x+iq_2y}\lan 0|\T \{ j_\mu(x)j_\nu(y)\} |\pi^0(q)\rrr =
$$
\be
= \epsilon_{\mu\nu\alpha\beta}q_1^\alpha q_2^\beta {\cal F}_\pi(q^2,q_1^2,q_2^2)
\ee
one gets at the leading order in constant external field:
\be
\lan 0|  j_\mu(-q) |\pi^0(q)\rrr_{F} = ie  q^\rho {\tilde F}_{\rho\mu} \> {\cal F}_\pi(q^2,q^2,0).
\label{fk2}
\ee
The expressions for $\eta$ and $\eta'$ contributions are completely analogous with the replacement of ${\cal F}_\pi$ by ${\cal F}_\eta$ and ${\cal F}_{\eta'}$.

Thus the $q^2$-dependence of polarization operator in external field
is determined in this approximation by the form-factors ${\cal
F}_\phi (q^2,q^2,0)$ with one on-shell  leg (corresponding to
external field vertex). These form-factors are essentially
nonperturbative QCD objects. Let us remind that on-shell (i.e. at
the point  ${\cal F}_\phi(m_\phi^2,0,0)$) they are fixed by triangle
anomaly, for example for pion: \be {\cal F}_\pi(m_\pi^2,0,0) =
-\frac{N_c}{12\pi^2 F_\pi} \ee Another important case is large $q^2 \to \infty$
limit where one has (for chiral fermions) $ {\cal F}_\phi(q^2,q^2,0) \to \chi_F F_\pi / 3 $ where
$\chi_F$ is QCD quark condensate magnetic susceptibility, defined by
$ \lan 0| \bar q \sigma_{\mu\nu} q |0\rrr_F = e_q \chi_F \lan \bar q
q \rrr F_{\mu\nu} $. Different approximation schemes valid at
intermediate momenta are discussed in the literature (see, e.g. \cite{gorsky}).

Having written the field-dependent matrix element (\ref{fk2}) one is able to express the invariant function $\tilde \pi^{(F)}(q^2)$ as follows:
\be
\tilde{\pi}^{(F)}(q^2) = \sum\limits_{\phi=\pi,\eta,\eta'} \frac{|{\cal F}_\phi(q^2,q^2,0)|^2}{q^2-m_\phi^2}
\ee
From the point of view of expression (\ref{as}) the dominant contribution to asymmetry is this phase comes from the lightest degree of freedom, i.e. massless in the chiral limit pion (to be more precise, we assume the limit $m_\pi R \ll 1$). Choosing for concretness Gaussian boundary condition (i.e. introducing the factors $\exp(-q_i^2R^2/2)$ into (\ref{as}) one obtains
\be
\lan q_V^2 \rrr = \gamma \left(\frac{eB}{F_\pi}\right)^2 T R^3
\ee
where the numerical factor $\gamma = 1.6 \cdot 10^{-4}$ is of course specific for this boundary choice. Certainly the result trivially follows from dimensional considerations. We see $\lan q_V^2 \rrr \ll 1 $ for phenomenologically reasonable choice of parameters. Contributions of mass gapped states bring additional suppression (and, in particular, break $\sim R^3$ scaling).

As the second example we consider free fermions in strong field limit. This regime would correspond to deconfinement phase where proper dynamical degrees of freedom are quarks and gluons with perturbatively weak interaction between each other. To compute polarization operator under external conditions in perturbation theory one usually makes use of Schwinger proper-time technique and there is extensive literature on the subject \cite{tsai,urrutia,adler,chodos,persson} where different kinds of external backgrounds were studied. The polarization operator in constant magnetic field and at nonzero temperature was calculated in \cite{alexander} in imaginary time formalism. Our aim here is to put these results in a charge fluctuations asymmetry prospective. For the reader's convenience we reproduce the explicit one-loop expressions
 for polarization operator $\Pi_{||}$ given by \cite{alexander} in Appendix of the present paper.

It is convenient to present the Euclidean polarization operator in the following form \be
\Pi_{\mu\nu}(q_\bot,q_3,n) = {\bf \sum} A_{\mu\nu}(q)\>e^{-\phi(q)} + Q_{\mu\nu}(q)
\ee
where the sum includes integration over proper-times and summation over Matsubara frequencies (see expression (\ref{sum}) in the Appendix), the functions $A_{\mu\nu}[q]$ polynomially depend on momenta components $q$, and the universal Euclidean phase $\phi(q)$ is given by expression (\ref{phase}). The contact terms $Q_{\mu\nu}(q)$ have no sensitivity to infrared parameters (like temperature or external field) and provide correct limit of $\Pi_{\mu\nu}$ at vanishing background.

One can notice that  ${\tilde \pi}^{(F)}$ can be simply related to the polarization operator components. Namely, solving the system of three
linear equations (\ref{expansion2}) for the choices $(\mu\nu)= 44, 33$ and $34$ one finds all three invariant form-factors, including ${\tilde \pi}^{(F)}$:
\be
B^2 {\tilde \pi}^{(F)} = -\frac{q_3q_4\Pi_{44} + (q_\bot^2 + q_3^2)\Pi_{34}}{q_\bot^2 q_3 q_4} \label{pitf}\ee
where $q_\bot^2 = q_1^2 + q_2^2$ and $q_4 \equiv $ $\omega_n =$ $ 2\pi T n$.

Thus the chiral magnetic form-factor is a nontrivial linear
combination of $\Pi_{34}$ and $\Pi_{44}$.
First of all we are to check that at $B\to 0$ the r.h.s. of (\ref{pitf}) vanishes. This is obvious at zero temperature since in this case there is the only  tensor structure given by $\Psi^{(1)}_{\mu\nu}$ and
\be
q_3 q_4 \Psi^{(1)}_{44} + (q_\bot^2 + q_3^2)\Psi^{(1)}_{34} = 0
\ee
For temperature-dependent parts it is rather nontrivial, the proof that this is indeed the case can be found in Appendix.

The explicit expression for  ${\tilde \pi}^{(F)} $ looks especially simple in small $T$ regime. It reads
\begin{widetext}
\be
{\tilde \pi}^{(F)} = -\frac{1}{(4 \pi)^2} \frac{1}{eB} \int_\epsilon^\infty du \int_{-1}^{+1} dv \>\left( (1-v^2) \coth {\bar u} + f_\bot({\bar u},v) \right)\exp(-\phi^{(0)})
\label{p}
\ee
\end{widetext}
where ${\bar u} = ueB$ and the functions $\phi^{(0)}$ and  $f_\bot({\bar u},v)$ are given in the Appendix. Notice that such form-factor was discussed in a different context in \cite{bk}.

In the weak field limit one has
\be
\lim\limits_{B\to 0} \> {\tilde \pi}^{(F)} = \frac{1}{6 {\pi}^2} \int_{-1}^1 dv \frac{(1-v^2)(3-v^2)}{\left(4m^2 + (1-v^2) q^2\right)^2}
\label{p4}
\ee
In the strong field limit (still at small $T$) the situation becomes more interesting - form-factor ${\tilde \pi}^{(F)} $ provides dominant contribution to the polarization operator:
$$
\Pi_{44} \to q_3^2 (eB)^2 {\tilde \pi}^{(F)} \to
$$
\be \to  -\frac{eB}{4\pi^2} \>
e^{-\frac{q_\bot^2}{2|eB|}} \>\int_{-1}^1 dv  \frac{(1-v^2)q_3^2}{4m^2 + (1-v^2)q_3^2}
\label{poj}
\ee
up to the terms ${\cal O}\left( q_\bot^2/eB \right)$. One can say that all asymmetry of charge fluctuations is due to CME-like formfactor in this limit.

We see another interesting effect - in the chiral limit (\ref{poj}) does not depend on $q_3$ at all, while the dependence on $q_\bot$ is suppressed by the field $B$. On the other hand, the essence of the asymmetry of interest is just different dependence of the polarization operator
on different components of momentum. Since the polarization operator itself linearly rise with $B$ for strong field it is nor a priori clear
which effect is to win. Detailed calculation shows that in fact they balance each other  and the asymmetry (\ref{as}) is not asymptotically rising with $B$ - there is an effect of saturation. It is reasonable to separate different regimes depending
on ratios between basic parameters such as $B$, $m$, $T$ and $R$ where the latter one stays for the typical 3-dimensional size of the volume $V_3$.
For two light flavors one can safely neglect quark masses $m$. Three other parameters are in the ballpark of 100 MeV (for large fireball one can think of phenomenologically realistic $eBR^2 = 5\div 10$). Without intention to cook up numerical factors but just to get feeling of the numbers,
pluging (\ref{poj}) into (\ref{as}) we get
\be
\lan q_V^2 \rrr = {\gamma}' \cdot R T
\ee
where again the numerical factor ${\gamma}' =4.1\cdot 10^{-2}$ corresponds to Gaussian boundary shape. Thus for asymptotically large $B$ one reaches
"kinematical limit" for the asymmetry in our picture, despite numerically it is still very small.

\section{Conclusions}

We have discussed three possible ways to study quantum physics behind chiral magnetic effect and electric charge fluctuation asymmetry
observed in heavy ion collisions. For all approaches the importance of scale separation is stressed - there should by hierarchy of dynamical scales
characterizing the life of quark-gluon phase after the collision and intrinsic QCD scales (perhaps field/temperature shifted)
characterizing the nonabelian topological charge fluctuation pattern. The physical essence of CME as we tried to present it here is that the quark-gluon medium plays the role of a measuring device with respect to the topological QCD vacuum with the final particles electric charge asymmetry as an outcome.
This is most clearly illustrated by the expression (\ref{fincur}).

The third approach we have considered, i.e. the analysis of P-odd $\times$ P-odd contributions to P-even observables, is somewhat different because it
provides nonzero results even for free fermions in magnetic field, i.e. without any "topological origin". We believe that this can be considered as a particular case of CME as well. Just nonzero matrix element of the vector current between vacuum and $J^{-+}$ states in external magnetic field leads to asymmetric charge/current pattern as if there is fluctuating vector current collinear to ${\bf B}$. Of course the detailed picture depends on the actual quantum dynamics of these $J^{-+}$ degrees of freedom, and we have shown that indeed it is strongly suppressed in the confinement phase. Nevertheless we find it legitimate to interpret this dynamics using  the same CME-like language
since namely this anomaly-driven vector-axial correlation is at the heart of the effect, while the concrete way of life of the axial degrees of freedom (distribution function for $\mu_5$ in the standard CME analysis) is of secondary importance.

We have left without attention all aspects of temperature dynamics in this paper. Despite no drastic qualitative effects are expected it is interesting to study the asymmetry in the whole parameter space spanned by ($B, T, m, R$). This could clearly have phenomenological applications to heavy ion collision physics whose understanding is the main challenge for modern QCD.

\acknowledgments

The authors acknowledge useful discussions with P.Buividovich, A.Gorsky, D.Harzeev, A.Kaidalov, M.Polikarpov, Yu.Simonov and V.Zakharov.
The work of one of the authors (V.O.) is supported by the "Dynasty" foundation personal grant.

\appendix*
\section{}

For the reader's convenience we present explicit expressions for
one-loop polarization operator as computed in \cite{alexander}. It
reads
\begin{widetext}
\be \Pi_{\mu\nu}({\bf q}, q_4) = -T\int\frac{d {\bf p}}{(2\pi)^3}
\sum\limits_{l=-\infty}^{\infty} {\mbox Tr}\left\{\gamma_{\mu}
S_l({\bf p})\gamma_{\nu}S_{l-n}({\bf p} - {\bf q})\right\} +
Q_{\mu\nu}(q) \ee
\end{widetext}
where $S_l({\bf p})$ is fermion propagator in external constant
magnetic field and $Q_{\mu\nu}(q)$ is the "contact term" needed to
cancel ultraviolet divergencies. It has no dependence on soft
backgrounds like temperature or external field. The sum goes over
fermionic Matsubara frequencies ${\hat \omega}_l = (2l+1)\pi T$,
while the bosonic one is $q_4 = \omega_n = 2n\pi T$.

Thus the general structure of Euclidean polarization operator is
given by \be \Pi_{\mu\nu}(q_\bot,q_3,n) = {\bf \sum}
A_{\mu\nu}(q)\>e^{-\phi(q)} + Q_{\mu\nu}(q) \ee The sum is given by
the following expression: \be {\bf \sum} = \frac{T
}{4\pi} \frac{eB}{\sqrt{\pi}} \int_\epsilon^\infty du \sqrt{u}
\int_{-1}^1 dv \sum\limits_{l=-\infty}^\infty \label{sum} \ee
The universal phase $\phi(q)$ has the form
$$
 \phi(q) = \phi^{(0)}(q) + u W_l^2 = \frac{q_\bot^2}{eB} \frac{\cosh {\bar u} - \cosh {\bar
u} v}{2 \sinh {\bar u}} + $$ \be + u\left[m^2 + W_l^2
+\frac{1-v^2}{4} (q_4^2 + q_3^2)\right] \label{phase} \ee where $W_l
= {\hat\omega}_l - \frac{1-v}{2}\omega_n$ and ${\bar u} = ueB$. The
contact term is given by \be Q_{\mu\nu} = \frac{1}{12\pi^2}
\int\limits_\epsilon^\infty \frac{du}{u} e^{-um^2}
\left(q^2\delta_{\mu\nu} - q_\mu q_\nu\right) \ee The function
$A_{\mu\nu}(q)$ polynomially depends on momenta components
$q$ and for 3,4 components reads $$ A_{44}(q) = \coth {\bar
u}\left(\frac{1}{u} - 2W_l^2 + q_4 v W_l -
\frac{1-v^2}{2}q_3^2\right) +
$$
\be + \frac{q_\bot^2}{2}f_\bot({\bar u},v) \label{a44} \ee
\be
A_{34}(q)=q_3\left[vW_l + \frac{1-v^2}{2} q_4\right]\coth{\bar u}
\label{a33} \ee
$$ A_{33}(q) = - \coth {\bar
u} \left[ q_4^2 \frac{1-v^2}{2}   +  q_4 v W_l \right] + $$
\be + \frac{q_\bot^2}{2}f_\bot({\bar u},v) \label{a34} \ee
where
\be
f_\bot({\bar u},v) = \frac{v\coth{\bar u} \sinh {\bar u} v - \cosh {\bar u}v
}{ \sinh {\bar u}}
\label{fbot}
\ee

To get zero temperature limit of the above expressions Poisson summation formula is useful
\be
\sum\limits_{l=-\infty}^{\infty} e^{-a(l-z)^2} = \left( \frac{\pi}{a} \right)^{1/2} \> \sum\limits_{k=-\infty}^{\infty} e^{-\frac{\pi^2k^2}{a} - 2\pi i z k }
\ee
In particular one gets
\be
\lim\limits_{T\to 0} T \sum\limits_{l=-\infty}^{\infty} e^{-uW_l^2} = \frac{1}{2\sqrt{u\pi}}
\ee

Another necessary demonstration of self-consistency is a proof of vanishing of $B^2 {\tilde\pi}^{(F)}$ defined by (\ref{pitf}) at $B=0$ for any $T$.
One has, at $B\to 0$
$$
q_3q_4 A_{44} + {\bf q}^2A_{34} =\frac{q_3}{u}\left(q_4\left(\frac{1}{u} - 2W_l^2\right) + vW_l{q}^2\right) =
$$
\be
= \frac{q_3}{2u^2} \frac{d}{dv} \left( W_l e^{-u(W_l^2 + \frac{1-v^2}{4}q^2)} \right)
\ee
It is easy to check that the latter expression gives zero result when integrated from -1 to 1.

\end{document}